\documentclass[12pt]{article}

\usepackage[T1]{fontenc}
\usepackage[T2A]{fontenc}
\usepackage[utf8]{inputenc}
\usepackage[english]{babel}
\usepackage{courier}
\usepackage{graphicx}
\usepackage{float}
\usepackage{amsthm}
\usepackage{amssymb}
\usepackage{concrete}
\usepackage{geometry}
\usepackage{caption}
\usepackage{subcaption}
\usepackage{algorithm}
\usepackage{algorithmic}
\usepackage{tikz}
\usetikzlibrary{decorations.pathreplacing}
\usepackage{wrapfig}
\usepackage{multirow}
\usepackage{multicol}
\usepackage{authblk}

\usepackage{amsmath}

\theoremstyle{plain}

\theoremstyle{definition}

\theoremstyle{remark}

\newcommand{\rank}{\mathrm{rank}\kern 2pt}

\floatname{algorithm}{Procedure}

\renewcommand{\geq}{\geqslant}
\renewcommand{\leq}{\leqslant}

\newcommand{\Fig}[1]{figure \ref{#1}}
\newcommand{\Figl}[1]{Figure \ref{#1}}
\newcommand{\Figls}[1]{Figures \ref{#1}}
\newcommand{\Figs}[1]{figures \ref{#1}}
\definecolor{green}{rgb}{0,0.7,0}

\begin{document}

\title{Collision  fragmentation of aggregates. The role of the interaction potential between comprising particles.
}
\author[1]{Alexander Osinsky}
\author[1,2]{Nikolai Brilliantov}
\affil[1]{Skolkovo Institute of Science and Technology, Moscow, Russia}
\affil[2]{Department of Mathematics, University of Leicester, Leicester LE1 7RH, United Kingdom}

\maketitle

\begin{abstract}
We investigate  disruptive collisions of aggregates comprised of particles with different interaction potentials. We study Lennard-Jones (L-J), Tersoff, modified L-J potential and the one associated with Johnson-Kendall-Roberts (JKR) model. These refer  to short, middle and long-ranged inter-particle potentials and  describe both inter-atomic interactions and interactions of macroscopic adhesive bodies. We perform comprehensive molecular dynamics simulations  and observe  for all four potentials power-law dependencies for the size distribution of collision fragments and for their size-velocity correlation. We introduce a new fragmentation characteristic -- the  shattering degree $S$, quantifying the fraction of monomers in debris and reveal its universal behavior. Namely, we demonstrate  that for all potentials, $1-S$ is described by a universal function of the impact velocity. Using the above results, we perform the impact classification and construct the respective collision phase diagram. Finally, we present a qualitative theory that explains the observed scaling behavior. 
\end{abstract}

\section{Introduction}\label{int-sec}
 Fragmentation phenomena are ubiquitous in nature and industrial processes. They occur at different time and space scales, and numerous examples may be mentioned. Fragmentation is studied in diverse research areas, ranging from geophysics~\cite{Grady1980,Turcotte1997} and astrophysics~\cite{Michel2003,Nakamura2008} to  surface physics \cite{Sokolov:1996,Sokolov:1997}, engineering~\cite{Thornton1996}, material \cite{Lankford1991} and  military science~\cite{Mott1943,Grady2006}. A surprising common feature of the fragmentation phenomenon is a power-law distribution of the mass of debris. It is observed for various spatial scales and nature of colliding objects -- collisions of heavy ions with a target or asteroids impact -- such distinct systems demonstrate  similar collision output, see 
e.g.~\cite{Herrmann1990,Nakamura-e-Fujiwara-1991,Kadono:1997,Giblin-etal-1998,Arakawa1999,Ryan-2000,Herrmann-etal-2006,Kun-etal-2006,Guettler2010,wall}. This supports a popular conjecture about a common physical principle  inherent to all fragmentation phenomena \cite{Kun1999,Astrom:2006}. 

Fragmentation has been extensively studied experimentally, e.g. \cite{exp, exp2}, theoretically and numerically, e.g. \cite{Herrmann1990,Astrom2004,Grady2009,Bouchaud-etal-1993, Galybin-e-Dyskin-2004,Paolicchi-etal-1996,Chang-etal-2002a,Kun1999,Astrom:2006,Spahn2014,ZhangPRL,plates,basis-conc,wall}. Many different approaches, including continuum mechanics, micro-structural mechanics, statistical methods and stochastic simulations have been applied. The main focus of all these studies was, however, on the mass/size  distribution of the debris. Different shapes of colliding bodies, such as, e.g. plates  \cite{plates}, or clusters obtained by ballistic cluster-cluster aggregation \cite{basis-conc} were investigated. Although  many studies reported a power-law for the fragment  mass distribution, $P(m) \sim m^{-\alpha}$, the exponent $\alpha$ was not universal \cite{basis-met,basis-sph,basis-sph2}. Moreover, it depended on the impact velocity   \cite{basis-sph}. Still, the important characteristics of disruptive collisions -- the velocity distribution of debris and its correlation with the size of the fragments have not been explored in detail yet; such correlations have been analyzed only in \cite{basis}. The authors have found  numerically that $v \sim m^{-1/3}$,  where $v$ is the velocity of a fragment and  $m$ is its mass.  Up to now, there have been no systematic studies of the role of the interaction potential between particles comprising the colliding bodies on the characteristics of fragmentation processes. The goal of the present article is to analyze the impact of such a potential on the size distribution of fragments and the correlation between the size and velocity of debris. We considered four different interaction potentials, quite contrast in their properties: The common Lennard-Jones (L-J) potential, carbon Tersoff potential and the potential associated with Johnson-Kendall-Roberts (JKR) interactions  between macroscopic adhesive bodies. The first two potentials are atomic potentials  -- the L-J potential is a simplified one, while Tersoff is more realistic, reflecting  peculiarities of carbon atoms interactions. The third, JKR potential, may be used to describe aggregates comprised of macroscopic particles, like ice particles aggregates in astrophysics, see, e.g. \cite{PNAS,AnaGu2010}. Additionally, we studied a specially constructed Lennard-Jones-like  potential with an extended width of the potential well. This has been done to explore the role of the interaction range of a potential. Despite the dramatic difference of these potentials we observed a lot of similarity in fragmentation processes and even found some universal behavior for the characteristics of debris.  We also present a  phase diagram classifying the collision outcome for the above potentials and develop  a qualitative theory. 

The rest of the article is organized as follows. In the next Sec. II we describe the model, and in Sec. III we report the numerical results. In Sec. IV we present a qualitative theory and discuss the collisions classification. Finally, in Sec. V we summarize our findings. Some simulation detail and useful information are given in the Appendix. 

\section{The Model}\label{mod-sec}

For simplicity, we will consider head-on collisions of  spherical aggregates of equal size at zero temperature. They are comprised of the constituents -- monomers, which could be either atoms or macroscopic particles.  We analyze four potentials for the interacting monomers. By choosing the appropriate mass units, we set  for all of the studied potentials, the  mass of monomer $m_0$ to be one, $m_0=1$  . The first potential is the Lennard-Jones potential,
\begin{equation}
\label{LJ}
  U_{L-J} = 4 \varepsilon \left[ \left( \frac{\sigma}{r} \right)^{12} - \left( \frac{\sigma}{r} \right)^6 \right],  
\end{equation}
with the characteristic energy $\varepsilon$ and characteristic length $\sigma$. It gives the interaction force   $F_{L-J} = 24 \varepsilon \sigma^{-1} \left[ 2 ( \sigma/r)^{13} - ( \sigma/r)^{7} \right]$. For this potential we will use the re-scaled distance $r \to r / (2^{1/6} \sigma)$ and re-scaled time, 
$t \to  t\sqrt{12 \varepsilon /(2^{1/6} m_0 \sigma^2)}$; it yields the re-scaled force,
\begin{equation}
\label{LJ-red}
F_{L-J} = 1/{r}^{13} - 1/{r}^{7}.
\end{equation}
For notation simplicity we keep the same symbols for the re-scaled quantities  as for the initial ones. Another potential  is the Tersoff potential for carbon, with the parameters from \cite{ters1989}. Due to a lengthy expression for this potential, the respective detail as well as the according time and length units are given in the Appendix. Next we consider the potential associated with the JKR interactions. Here we present  the corresponding force acting between spherical particles of diameter $d$.  If the inter-center distance of two particles $r$ is smaller than $d$, they are squeezed and form a contact of radius $a =a(r)$. Then the force is determined by the relations \cite{Johnson1971,Brilliantov2007}: 

\begin{equation}\label{jkr-force}
  \begin{aligned}
  F_{JKR} \left( a \left( r \right) \right) & = 
  \left( 4 a^3 - 3 \varkappa a^{3/2}d \right)/(d D), \\
  d - r & = (4 a^2 - 2 \varkappa a^{1/2} d)/d. 
  \end{aligned}
\end{equation}
Here $\varkappa = \sqrt{\frac23 \pi \gamma D}$, with the adhesion coefficient  $\gamma$,  which is twice the surface free energy per unit area of a solid in vacuum. $D = \frac32 (1 - \nu^2) / Y$ is the elastic constant, where $Y$ is the Young's modulus, and $\nu$ is the Poisson ratio. For $r>d$ the inter-particle force is zero  for approaching particles, and is negative for moving-away particles, for the short distances $|r-d| \ll d$. Hence it is a  very short-ranged potential. The natural unit length for this potential is $d$, that is, we re-scale the length as $r \to r/d$ and time as $t \to t \sqrt{d D / (m_0 D_0)}$, keeping again the same notations for the  re-scaled variables. The dimensionless parameter $D_0$ is needed to adjust the time scale for collisions of JKR clusters, to be comparable with that of L-J clusters.  The re-scaling yields, 
\begin{equation}\label{jkr-force2}
  \begin{aligned}
  F_{JKR} \left( a \left( r \right) \right) & = 
  \left( 4 a^3 - 3 \varkappa a^{3/2}\right)/D_0, \\
  1 - r & = 4 a^2 - 2 \varkappa a^{1/2}, 
  \end{aligned}
\end{equation}
with the  dimensionless parameter  $\varkappa$, that  determines the equilibrium distance. It may vary in a wide range, from the one limit for soft adhesive particles, to another limit for hard particles, with  weak adhesion. Since our goal is  to compare  fragmentation of aggregates with  different potentials, we choose the following parameters (in the reduced units): $D_0 = 1.12 \cdot 10^{-5}$ and $\varkappa = 5.2 \cdot 10^{-3}$; this  results in the disruptive collisions at approximately  same (reduced) impact speeds  as for the L-J aggregates. The equilibrium distance between monomers for these  parameters (when $F_{JKR} \left( a \left( r \right) \right)=0$), is equal to $r \approx 0.9992$. The minimal distance between particle centers in our collisional simulations was never less than $0.95$. The values closer to one, require a smaller time step, increasing the simulation cost. 

\begin{wrapfigure}{R}{0.5\textwidth}
\centering
\includegraphics[width=0.49\columnwidth]{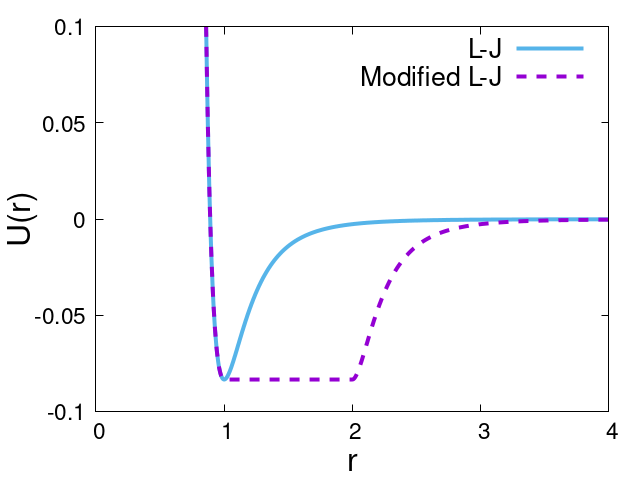}
\caption{Lennard-Jones and modified Lennard-Jones potential as used in the numerical experiments. The energy and length are given in the reduced units, $\varepsilon$ and $\sigma$ respectively, see Eq. (\ref{LJ}).
}
\label{LJ-mod-pot-fig}
\end{wrapfigure}

Finally, we analyze the modified Lennard-Jones potential,  obtained from the L-J potential (\ref{LJ}) by adding a flat plateau of the length 1 at its  bottom, see \Fig{LJ-mod-pot-fig}. The number of neighbors with which a monomer interacts for the modified L-J potential significantly exceeds that for the standard LJ potential. Therefore we will call this potential "long-ranged". Note  that commonly, the term "long-range"  is applied for the potentials, which  scale as $\sim 1/r^b$, with $b\leq 3$; here we use this term for convenience.  By studying the modified potential,  we explore the role of the interaction range of a potential, between comprising aggregates particles,  on the characteristics of the fragmentation process. The re-scaled length and time for the modified L-J potential are the same as for the standard one. 

\section{Simulation  results}\label{res-sec}
\subsection{Simulation detail}

The colliding spherical aggregates were prepared by fixing their  radius $R$ and placing monomers inside the sphere  with the closest (FCC - face-centered cubic) packing. For the aggregates of $R = 5$ this resulted in $757$ particles (of unit diameter) inside  each sphere, and for $R=10$ there were $5947$ particles. To make the packing more random, we set the average square  velocities of the monomers to $\left\langle \vec v^2 \right\rangle = 0.1$ and then slowly cool the system down  to zero temperature $T = 0$; additionally, clusters were randomly rotated. The systems with the Tersoff potential were simulated with the use of LAMMPS \cite{lammps}, where this potential is built in. For the aggregates with  the Tersoff potential, we started with the  natural diamond packing, instead of FCC; for $R=2.87$ (in lattice units), clusters of $801$ monomers have been generated.

\begin{wrapfigure}{R}{0.5\textwidth}
\centering
\includegraphics[width=0.5\columnwidth]{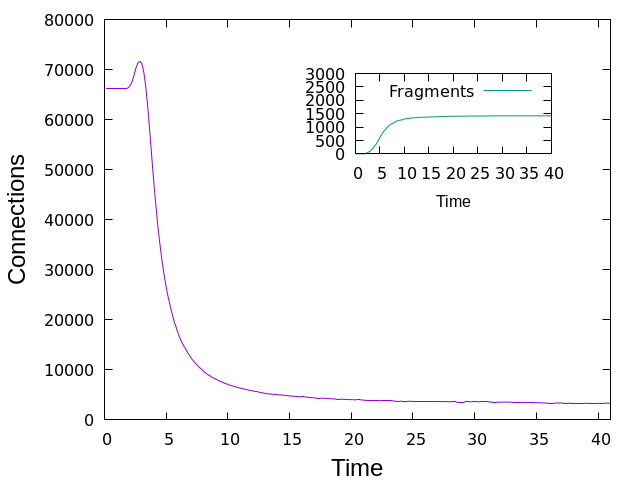}
\caption{The time dependence of the total number of closely connected pairs of particles (pairs within the distance $2$) for Lennard-Jones potential. Inset: The number of fragments as a function of time.}
\label{conn-fig}
\end{wrapfigure}

The numerical algorithm is detailed in the Appendix. Here we only mention that we used the same time step for all simulation parameters, which was chosen to guarantee that the  difference of the system energy at the beginning and the end of simulations did not exceed  $0.5\%$. For most of the simulations we performed from 1000 to about 6600 runs  for each set of parameters. The particular number of runs was dictated by the need for sufficient statistics. 

We solved  the equations of motion for all constituents during the complete impact until the aggregates or debris move far away from each other. Namely, we continued the simulations until one of the particles left the box of size $200 \times 200 \times 200$, centered at the system center of mass. \Figl{conn-fig}, indicating  the time dependence of the  closely connected monomer pairs, proves that  by this time, the collision is long over.


After the end of an impact, we evaluated the number of fragments and their kinetic energy. To identify different fragments, one needs to introduce the connectedness criterion. The bonds for the JKR interactions were explicitly recalculated every time step, using the criterion of inter-particles force, $F_{JKR} \neq 0$, since some particles at the range $r > 1$ may be connected, while the other -- disconnected. The Tersoff potential includes an explicit cutoff (see the Appendix), which can be used to determine the presence of bonds. For the L-J and modified L-J potentials, we consider the constituents to be connected for $r \leq 2 $ and $r \leq 3$ respectively  (again in the reduced units). 

More simulation detail, including the time steps and application of the software, are given in the Appendix. 

\subsection{Mass and speed distribution of debris }

\begin{figure}[h!]
\center
\begin{subfigure}[t]{0.49\textwidth}
\centering
\includegraphics[width=\columnwidth]{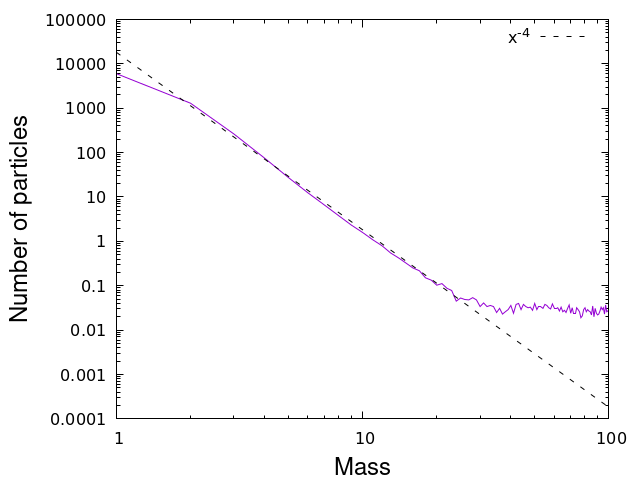}
\caption{Johnson-Kendall-Roberts (JKR) \\ interactions,  $R = 10$, $v_{\rm imp} = 2$, \\ averaging is over 1000 collisions.}
\label{fig:conc-jkr}
\vspace*{3mm}
\end{subfigure}
\begin{subfigure}[t]{0.49\textwidth}
\centering
\includegraphics[width=\columnwidth]{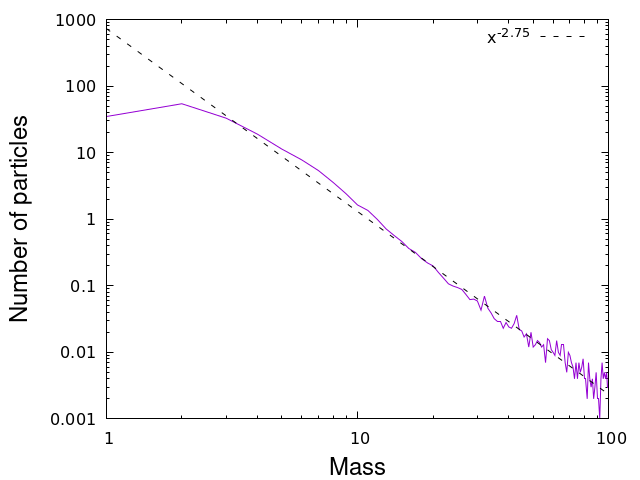}
\caption{Tersoff carbon potential, \\ $R = 2.87$ lattice units, $v_{\rm imp} = 200$, \\ averaging is over 1000 collisions.}
\vspace*{3mm}
\end{subfigure}
\begin{subfigure}[t]{0.49\textwidth}
\centering
\includegraphics[width=\columnwidth]{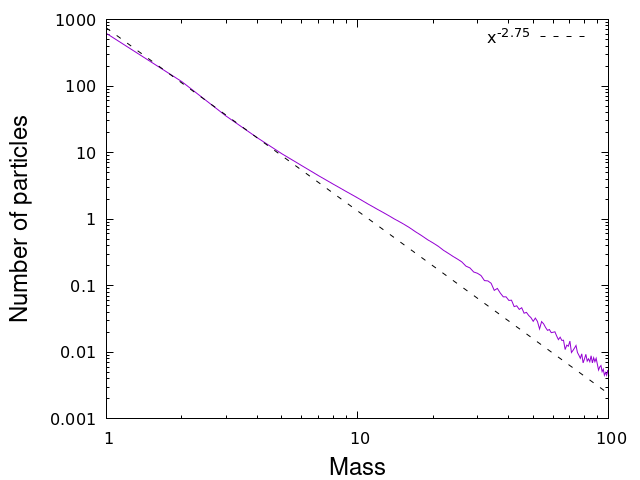}
\caption{Lennard-Jones potential, \\ $R = 5$, $v_{\rm imp} = 4$, \\ averaging is over 6635 collisions.}
\end{subfigure}
\begin{subfigure}[t]{0.49\textwidth}
\centering
\includegraphics[width=\columnwidth]{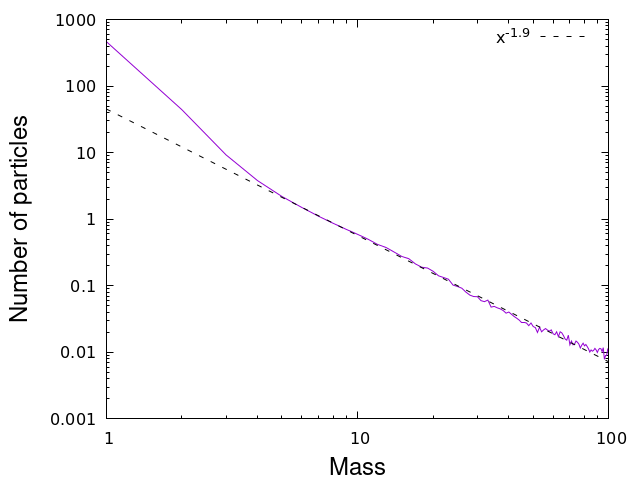}
\caption{Modified Lennard-Jones potential, \\ $R = 5$, $v_{\rm imp} = 5$, \\ averaging is over 6632 collisions.
}
\end{subfigure}
\caption{Fragment mass distribution -- the number of fragments of mass $m$ as a function of mass. Different potentials demonstrate the power-law dependence $\sim m^{-\alpha}$. The exponent $\alpha$ is not universal. }
\label{conc-fig}
\end{figure}

\Figl{conc-fig} shows the mass/size distributions of fragments for colliding aggregates with different  interaction  potentials between the comprising particles. In accordance with the previous experimental and numerical studies,  the mass distribution clearly demonstrates a power-law part for a significant range of debris size. Again, in accordance with the other studies, the exponent of the power-law, $P(m)\sim m^{-\alpha}$, is not universal. For the atomic, middle-ranged L-J and Tersoff potentials, $\alpha = 2.75$, which agrees with the simulation results of \cite{basis-sph2}, where $\alpha = 2.8 \pm 0.05$ has been reported. In contrast, for  the very short-ranged interactions, as the JKR interactions, a much steeper exponent of $\alpha =4$, with a strong prevalence of small debris, is observed. At the same time, for the relatively long-ranged, modified L-J potential  $\alpha = 1.9 $, which corresponds to much more abundant large fragments. This allows us to propose a conjecture -- the larger the interaction range of the inter-constituent potential, the larger the share of massive fragments in the size distribution of debris. And the opposite -- the shorter the interaction range, the larger the share of small fragments.

\begin{figure}[h!]
\center
\begin{subfigure}[t]{0.49\textwidth}
\centering
\includegraphics[width=\columnwidth]{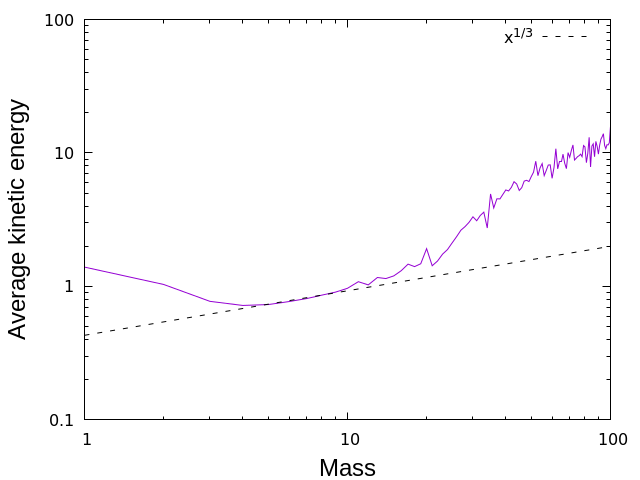}
\caption{Johnson-Kendall-Roberts (JKR) potential, \\ $R = 10$, $v_{\rm imp} = 2$, \\ averaging is over 1000 collisions.}
\vspace*{3mm}
\end{subfigure}
\begin{subfigure}[t]{0.49\textwidth}
\centering
\includegraphics[width=\columnwidth]{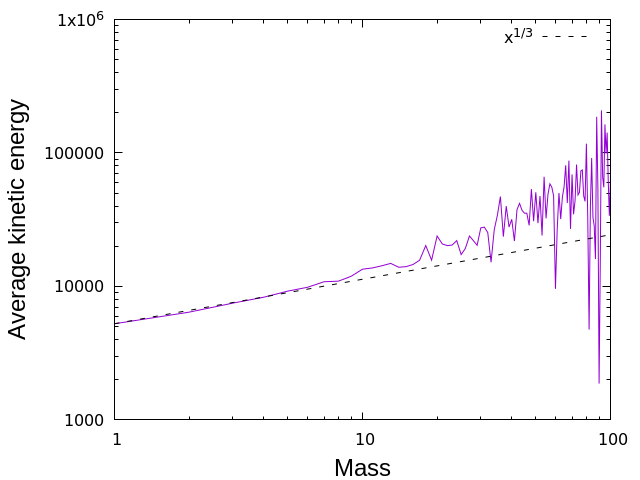}
\caption{Tersoff carbon potential, \\ $R = 2.87$ lattice units, $v_{\rm imp} = 200$, \\ averaging is over 500 collisions.}
\vspace*{3mm}
\end{subfigure}
\begin{subfigure}[t]{0.49\textwidth}
\centering
\includegraphics[width=\columnwidth]{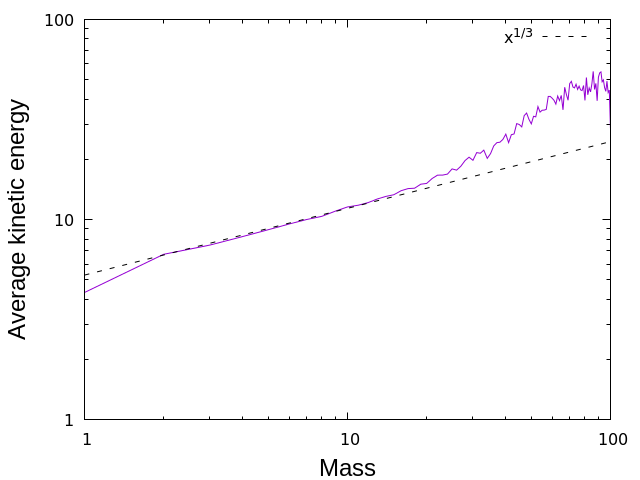}
\caption{Lennard-Jones potential, \\ $R = 5$, $v_{\rm imp} = 4$, \\ averaging is over 6635 collisions.}
\end{subfigure}
\begin{subfigure}[t]{0.49\textwidth}
\centering
\includegraphics[width=\columnwidth]{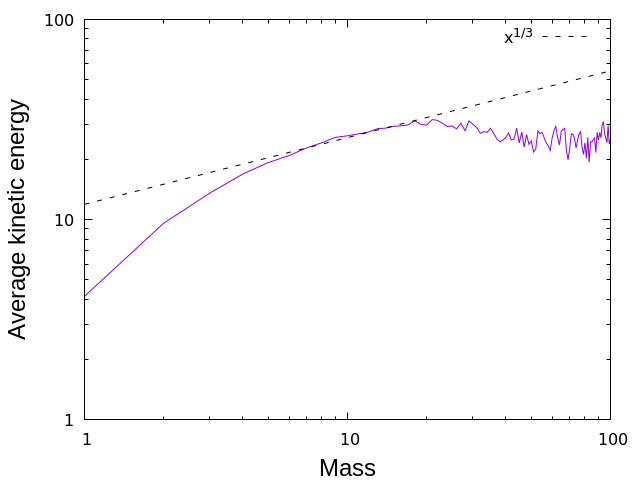}
\caption{Modified Lennard-Jones potential, \\ $R = 5$, $v_{\rm imp} = 5$, \\ averaging is over 6632 collisions.}
\end{subfigure}
\caption{Distribution of the kinetic energy of fragments. The very short-ranged (JKR) and long-ranged (modified L-J) interactions do not  demonstrate clearly the $m^{1/3}$- scaling for the fragment energy. }
\label{speed-fig}
\end{figure}

\Figl{speed-fig} shows the dependence of the kinetic energy  of fragments, $ E_{\rm kin, f}=m v^2/2$, on their mass $m$. As it follows from the figure, the dependence of $v \sim m^{-1/3}$, reported in \cite{basis} and corresponding to $E_{\rm kin, f} \sim m^{1/3}$,  is indeed  rather distinctly observed for the middle-ranged potentials --  L-J and Tersoff potential. It  could also be assumed, with some credibility, for a limited range of fragment mass,  for the short-range (JKR) and long-range (modified L-J) potential. In Sec. IV we present a qualitative theoretical justification for the  $v \sim m^{-1/3}$ dependence. As it follows from the theoretical analysis, one should not expect that the dependence holds for very small and very large fragments. 

\begin{figure}[h!]
\center
\begin{subfigure}[t]{0.2\textwidth}
\centering
\includegraphics[width=0.75\columnwidth]{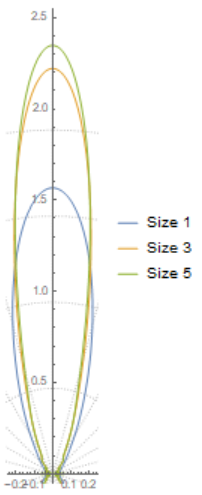}
\caption{Angular distribution for the velocity direction of fragments of size  1, 3, and 5.}
\label{fig:angle2a}
\end{subfigure}
\hfill
\begin{subfigure}[t]{0.35\textwidth}
\centering
\includegraphics[width=\columnwidth]{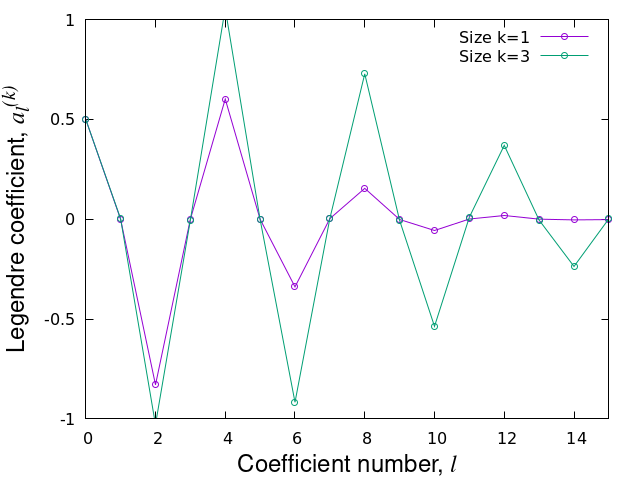}
\caption{The Legendre polynomials coefficients for monomers and triplets, $a^{(1)}_l $ and $a^{(3)}_l$,  as a function of $l$.}
\label{fig:angle2b}
\end{subfigure}
\hfill
\begin{subfigure}[t]{0.35\textwidth}
\centering
\includegraphics[width=\columnwidth]{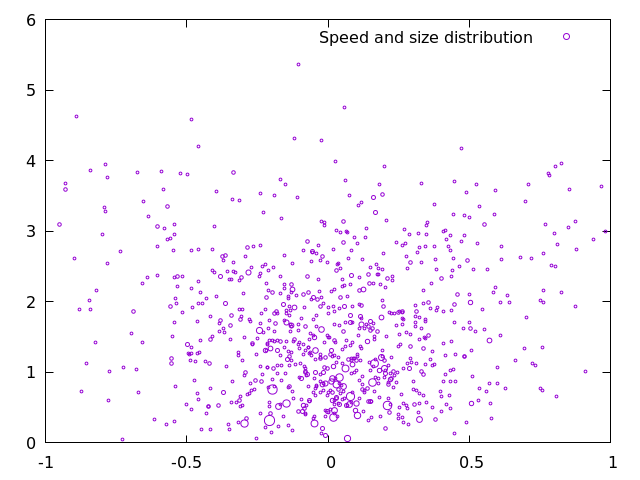}
\caption{Example of  the fragments velocity  distribution. Circle radii correspond to the fragment mass as $r \sim m^{1/3}$. The position of the circles on $x$ and $y$ axes shows, respectively,  the speed components parallel and perpendicular to the collision direction -- the inter-center line.}
\label{fig:angle2c}
\end{subfigure}
\caption{Angular distribution of the velocities direction of the Lennard-Jones fragments at $v_{\rm imp} = 4$.}
\label{angle-fig2}
\end{figure}

\begin{figure}[h!]
\center
\begin{subfigure}[t]{0.2\textwidth}
\centering
\includegraphics[width=\columnwidth]{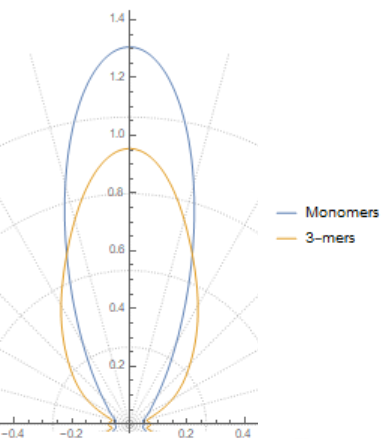}
\caption{Angular distribution of the velocity directions of the fragments -- monomers and 3-mers.}
\label{fig:anglea}
\end{subfigure}
\hfill
\begin{subfigure}[t]{0.35\textwidth}
\centering
\includegraphics[width=\columnwidth]{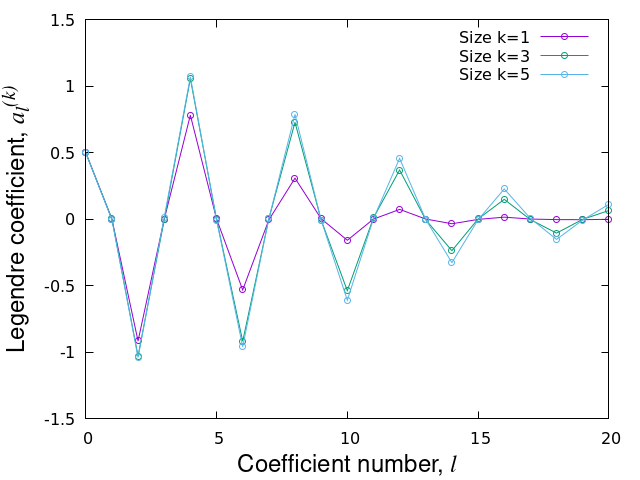}
\caption{The Legendre polynomials coefficients for monomers and fragments of size 1, 3, and 5, triplets, $a^{(1)}_l $, $a^{(3)}_l $ and $a^{(5)}_l$,  as a function of $l$.}
\label{fig:angleb}
\end{subfigure}
\hfill
\begin{subfigure}[t]{0.35\textwidth}
\centering
\includegraphics[width=\columnwidth]{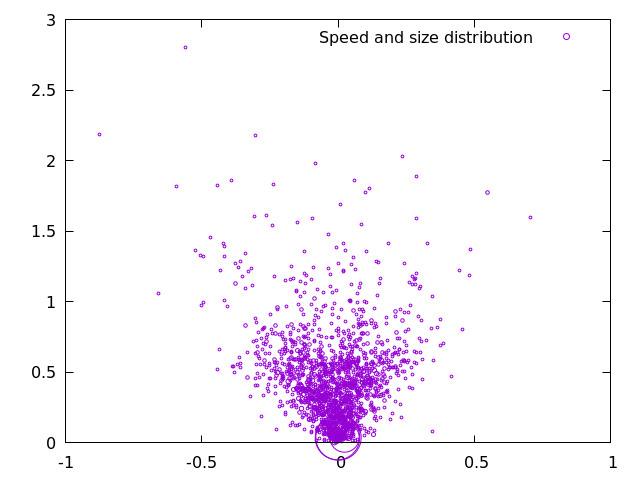}
\caption{Example of the fragments velocity  distribution. Circle radii correspond to the fragment mass as $r \sim m^{1/3}$. The position of the circles on $x$ and $y$ axes shows,  respectively,  the speed components parallel and perpendicular to the collision direction.}
\label{fig:anglec}
\end{subfigure}
\caption{Angular distribution of the velocity of JKR fragments at $v_{\rm imp} = 2$.}
\label{angle-fig}
\end{figure}

Next, we explored the angular distribution for  the directions of the fragment velocities. We focused  only on the middle-ranged (L-J) and short-ranged interactions (JKR). The results are presented in \Figs{angle-fig2} and \ref{angle-fig}. The impact velocities ($v_{\rm imp} = 4$ for L-J and $v_{\rm imp} = 2$ for JKR) were chosen to obtain a sufficient amount of debris. \Figls{fig:angle2a}  and \ref{fig:anglea} depict the distribution of the velocities direction for monomers and small clusters, up to size five, in the polar coordinates. As it may be seen from the plots, the direction of motion of small fragments  is sharply peaked around the plane, perpendicular to the inter-center line at the collision instant. To analyze the angular dependence in a more quantitative way we introduced  the angular distribution functions $F_k(\theta)$, so that $2 \pi F_k(\theta) \sin \theta \, d \theta$, gives the fraction of fragments of size $k$ with the velocity direction  within the angular interval $(\theta, \theta+ d \theta)$. Here the angle $\theta$ is the angle between the inter-center  vector $\vec{r}_{12}$ at a collision instant and the velocity of a fragment; we also take into account the azimuthal symmetry of the collision.  Due to the  collision mirror symmetry (recall that we study collisions of equal aggregates),  $F_k(\theta)=F_k(\pi-\theta)$, hence we do not distinguish between the angles $\theta$ and $\pi-\theta$ for the  function $F_k(\theta)$. It may be expanded in a  series of Legendre polynomials $P_l(\theta)$ as $F_k(\theta) = \sum_{l=0}^{\infty} a_l^{(k)} P_l({\theta})$, with the Legendre coefficients, 
\begin{equation}
\label{eq:al}
a_l^{(k)} = \frac{2l+1}{2} \int_{0}^{\pi} F_k(\theta) P_l({\theta}) \sin \theta \, d \theta.  
\end{equation}
Using the simulation data for $F_k(\theta)$, we computed the coefficients $a_l^{(k)}$ for small fragments, $k=1-5$, see \Figs{fig:angle2b}  and \ref{fig:angleb}. As expected, the odd coefficients $a_l^{(k)}$ almost vanish, reflecting the mirror symmetry, while the even coefficients rapidly decay with the order $l$.  The non-zero values of $a_l^{(k)}$ up to $l \simeq 15$ may be explained by the cusps at $\theta=0$ and  $\theta= \pi$. 

While the velocities of small fragments are mainly directed around the plane, normal to the inter-center line, the large fragments have much wider  distribution of their velocity directions. This is illustrated in \Figs{fig:angle2c}   and \ref{fig:anglec}. Here the position of a dot indicates the velocity direction of a fragment  in the inter-center and normal to it directions, while the size of the dot encodes the fragment size. One can observe that the middle-ranged interactions (L-J) result in a rather uniform distribution of the velocities directions; still the preferred direction is the one, perpendicular to the inter-center line. In contrast, 
for the short-ranged interactions (JKR) the velocities directions are much more focused around the perpendicular plane, resembling a collision of granular jets \cite{GranJetTho,Granjet2016}. 

\subsection{Impact of the collision velocity}
\begin{figure}[h!]
\center
\begin{subfigure}[t]{0.49\textwidth}
\centering
\includegraphics[width=\columnwidth]{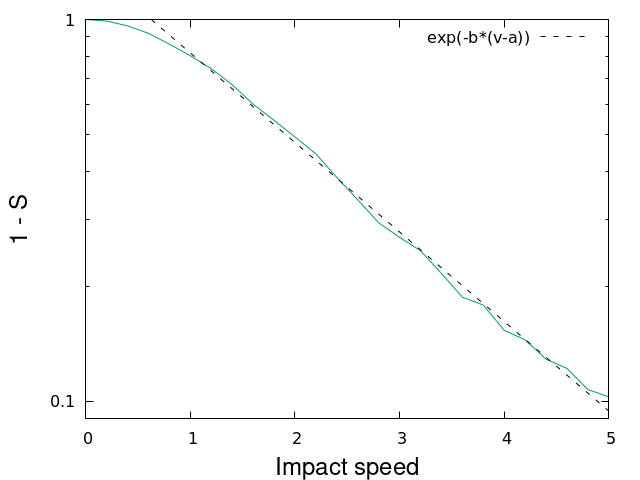}
\caption{Johnson-Kendall-Roberts (JKR) \\interactions,  $R = 10$, \\ averaging is over 10 collisions. \\ $a = 0.63 \pm 0.04, b = 0.540 \pm 0.008.$}
\label{fig-mona}
\vspace*{3mm}
\end{subfigure}
\begin{subfigure}[t]{0.49\textwidth}
\centering
\includegraphics[width=\columnwidth]{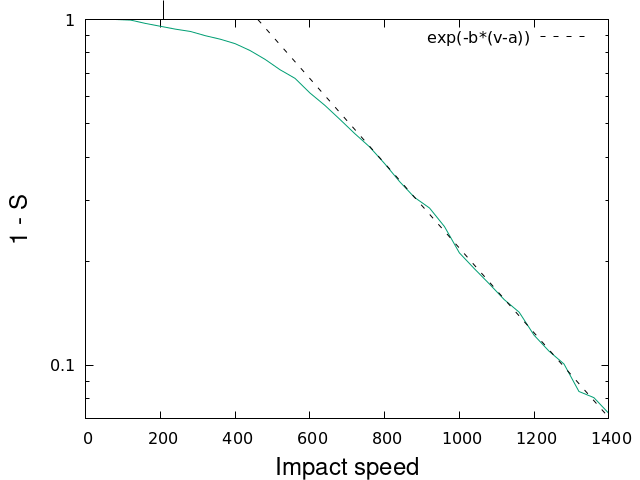}
\caption{Tersoff carbon potential,  $R = 2.87$ lattice units, \\ averaging is  over 10 collisions. \\ $a = 460 \pm 8, b = (2.82 \pm 0.05) \cdot 10^{-3}.$}
\vspace*{3mm}
\end{subfigure}
\begin{subfigure}[t]{0.49\textwidth}
\centering
\includegraphics[width=\columnwidth]{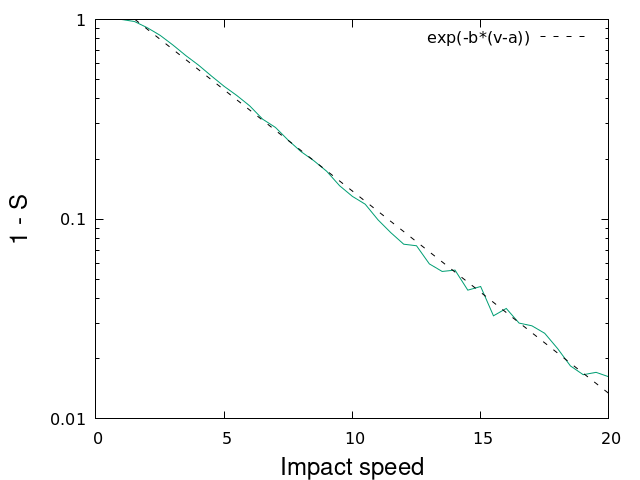}
\caption{Lennard-Jones potential,  $R = 5$, \\ averaging is  over 10 collisions. \\ $a = 1.52 \pm 0.12, b = 0.233 \pm 0.002.$}
\end{subfigure}
\begin{subfigure}[t]{0.49\textwidth}
\centering
\includegraphics[width=\columnwidth]{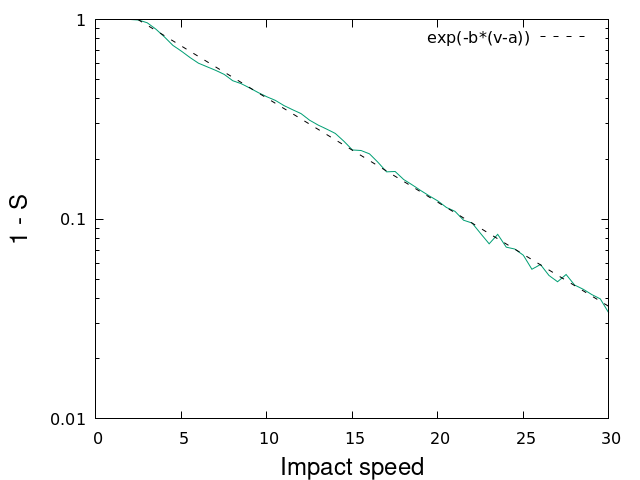}
\caption{Modified Lennard-Jones potential, \\ $R = 5$,  averaging is over 10 collisions. \\ $a = 2.45 \pm 0.11, b = 0.1201 \pm 0.008.$}
\end{subfigure}
\caption{Shattering degree, $1-S$, as the function of the impact velocity $v_{\rm imp}$ (see text for the definition of $S$) for different potentials. The dependence of $1-S$ on $v_{\rm imp}$ is universal.
}
\label{fig-mon}
\end{figure}

Now we analyze the impact of the collision energy, which is the total kinetic energy of colliding aggregates,  on the  characteristics of the fragmentation process. We observed  that the exponents of the power-law distributions discussed above persisted with the change of the collision energy, although the distributions for small and large mass altered. Indeed, very large collision energy  suffices for a complete disruption (shattering) of an aggregate into monomers. And oppositely, small collision energy results in the abundance of large fragments and the deficit of monomers.  The quantitative illustration of such  fragmentation behavior is given in \Fig{fig-mon}. Here the fraction of clusters (that is, fragments of size $k \geq 2$) among the debris is shown as a function of the impact velocity (recall, that two aggregates move with the velocity $v_{\rm imp}/2$ and suffer a head-on collision with the impact velocity $v = v_{\rm imp}$). Let us define a quantity $S$, which may be called a "shattering degree":  
\begin{equation}
\label{eq:S}
S = \frac{n_1}{2N}, 
\end{equation}
where $n_1$ is the number of monomers among the debris, $2N$ is the total number of monomers in both aggregates and $2N-n_1$ is the total number  of constituents belonging to any of the debris, but not monomers. The value of $S$ quantifies, how far is a collision outcome from shattering: $S=0$ implies no monomers at an impact, while $S=1$ -- a complete shattering. As expected,  with the increasing impact speed $S$ approaches one and $1-S$ -- zero, see \Fig{fig-mon}. Surprisingly, $1-S$ demonstrates a universal exponential behavior, 
\begin{equation}
\label{eq:S1} 
1-S= \exp \left[-b(v_{\rm imp}- v_{\rm imp, 0})\right],
\end{equation}
for $v_{\rm imp} \gtrsim  v_{\rm imp, 0}$, where $v_{\rm imp, 0}$ may be interpreted as a threshold impact speed separating two collision regimes (see the discussion below).  In the  next Sec. IV we discuss the theoretical justification  of this dependence.  

\begin{figure}[h!]
\center
\begin{subfigure}[t]{0.45\textwidth}
\centering
\includegraphics[width=\columnwidth]{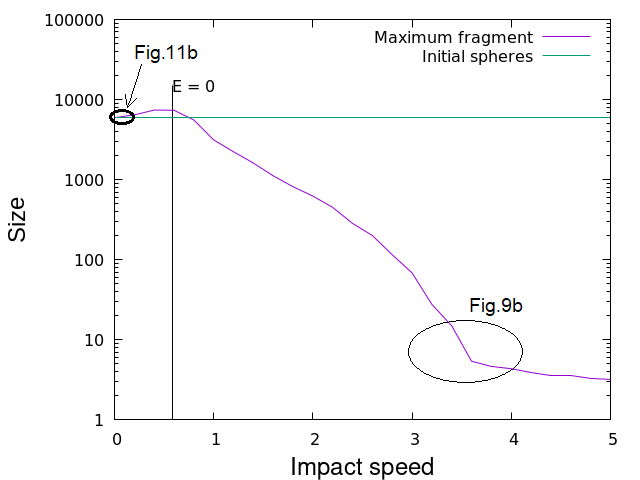}
\caption{Johnson-Kendall-Roberts (JKR) interactions, $N = 5947$ particles in each sphere, averaging is over 20 collisions. The circled areas are zoomed in \Fig{fig:examplesb} (small velocities) and \Fig{fig:moredetailb} (large velocities).  }
\label{JKR-maxmass-fig}
\vspace*{3mm}
\end{subfigure}
\hfill
\begin{subfigure}[t]{0.45\textwidth}
\centering
\includegraphics[width=\columnwidth]{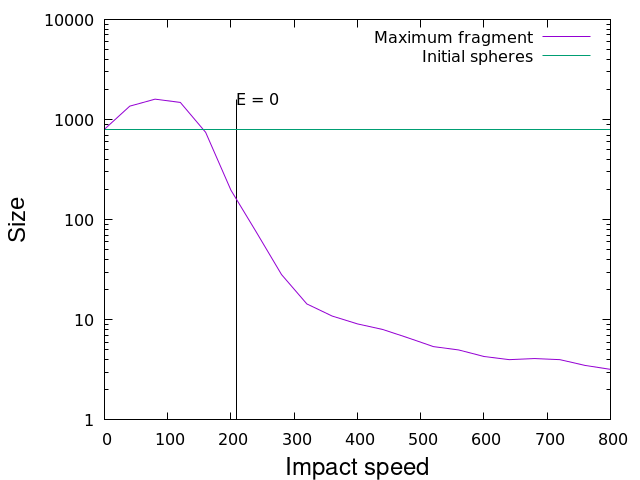}
\caption{Tersoff carbon potential, $N = 801$ particles in each sphere, averaging is over 10 collisions.}
\label{Ters-maxmass-fig}
\vspace*{3mm}
\end{subfigure}
\begin{subfigure}[t]{0.45\textwidth}
\centering
\includegraphics[width=\columnwidth]{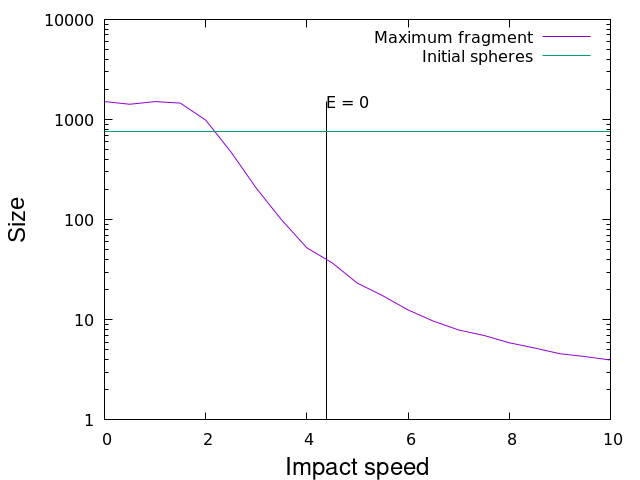}
\caption{Lennard-Jones potential, $N = 757$ particles in each sphere, averaging is over 100 collisions.}
\label{LJ-maxmass-fig}
\end{subfigure}
\hfill
\begin{subfigure}[t]{0.45\textwidth}
\centering
\includegraphics[width=\columnwidth]{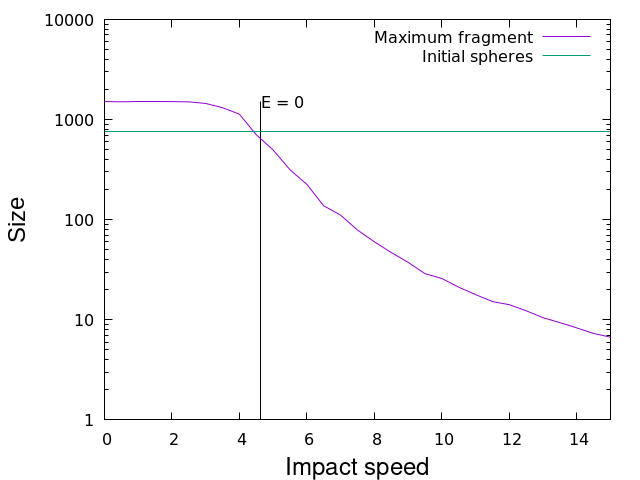}
\caption{Modified Lennard-Jones potential, $N = 757$ particles in each sphere, averaging is over 100 collisions.}
\label{LJm-maxmass-fig}
\end{subfigure}
\caption{The dependence of the maximum fragment size on the impact velocity $v_{\rm imp}$.  Each aggregate has $N$ particles, therefore maximum size of $N$ shows bouncing. The maximum size equals $2N$ indicates sticking without erosion. The line $E = 0$ shows the impact velocity leading to vanishing total energy after the collision. JKR potential has rather special behavior, detailed in \Fig{fig:moredetailb} and \Fig{fig:examplesb}.}
\label{maxmass-fig}
\end{figure}

\begin{figure}[h!]
\center
\begin{subfigure}[t]{0.45\textwidth}
\centering
\includegraphics[width=\columnwidth]{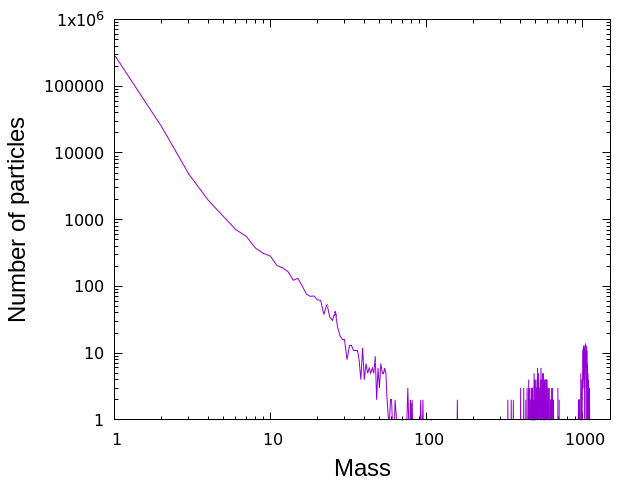}
\caption{Number of fragments for modified L-J potential at $v_{\rm imp} = 2$. Large ``fragments'' indi\-cate either bouncing or sticking with erosion.}
\label{fig:moredetaila}
\end{subfigure}
\hfill
\begin{subfigure}[t]{0.45\textwidth}
\centering
\includegraphics[width=\columnwidth]{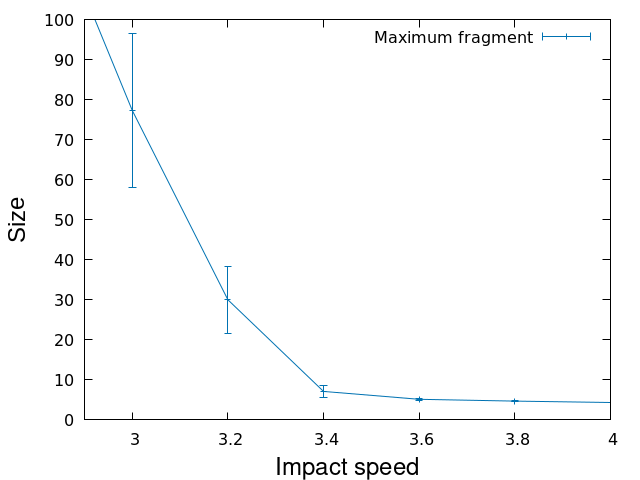}
\caption{The dependence of the maximum fragment size on the impact velocity for the JKR aggre\-gates, averaged over 20 collisions. Bars show the variance of maximum size for one collision. There is a sharp transition to the maximum size of about $5$ at $v_{\rm imp} \gtrsim 3.4$.}
\label{fig:moredetailb}
\end{subfigure}
\caption{Disruptive and mixed-type collisions -- a more detailed view. }
\label{fig:moredetail}
\end{figure}

\Figl{maxmass-fig} illustrates the dependence of the maximal fragment size on the impact velocity $v_{\rm imp}$. \Figl{fig:moredetail} supplements \Fig{maxmass-fig}, providing a more detailed view on the disruptive and mixed-type collisions. The impacts with small $v_{\rm imp}$ may result in  sticking. Hence, the maximal fragment size of $N$ corresponds to a bouncing collision, while the size of $2N$ refers to sticking. This is also illustrated in \Fig{fig:moredetaila} for mixed type collisions. \Figl{fig:moredetailb},  supplementing \Fig{JKR-maxmass-fig}, indicates a sharp transition to shattering for $v_{\rm imp}  \gtrsim 3.4$. The vertical lines in the plots show the impact velocity leading to vanishing total energy  $E$ at the end of the collision (recall that while the kinetic collision energy is positive, the potential energy of aggregates is negative, as the constituents are bound). As it follows from the figure, $E=0$ can be hardly used as a universal criterion that specifies the alternation  of a collision regime. 

\subsection{Classification of collisions }

Although  the main focus of the present study is the disruptive collisions, it would be worth considering this type of collisions within a more general framework, classifying all possible collision outcomes. This is challenging, as it is not always possible to observe a particular collision behavior in a pure form. \Figl{fig:moredetaila} illustrates this: in  multiple impact  realizations, there exist collisions, where apart from small fragments of about a few monomer sizes,  aggregates of size around $N$ and $2N$ are observed. The former corresponds to almost bouncing collisions, the latter -- to almost sticking one. One can call these processes "bouncing with erosion" and "sticking with erosion," respectively. The term "erosion" here  means the presence of two large fragments comparable in size  with  the initial aggregates (see also the discussion below).  Note that these collision outcomes occur for the same impact velocity of the same aggregates, demonstrating the stochastic nature of fragmentation. In simulations, the stochasticity emerges from slightly different initial conditions (the orientation of the aggregates) and numerical noise. In nature, it can arise from the randomness of the structural properties of aggregates. Said this, we consider  below the most probable collision outcome.

One  can define five main collision types: impacts with sticking, bounce, erosion, fragmentation and shattering. As we have already mentioned, the erosion may be accompanied by sticking or bounce. Although  somewhat subjective, we proposed the following definitions: 
(i) A purely sticking collision implies the formation of a joint aggregate from the colliding partners; (ii) a purely bouncing collision results in a change of the aggregates velocities only; (iii) in a collision with erosion, one or two large fragments of the size, at least half of the initial size,  remain -- a more quantitative definition follows below; (iv) a collision with fragmentation implies a lack of large debris, comparable in size with the initial dimension, see also below; (v) in a shattering collision only small debris appear; here we put the boundary -- all debris should have size  not larger than 5 monomers (about 0.3\% of the initial mass for $R=5$).  This is a relatively rough  classification. One can make  a more refined  classification, as proposed in \cite{lj-modes} --  the oblique impacts of L-J nanoclusters were also considered/classified there. Below we exemplify the above type of collisions for different potentials. 

\begin{figure}[ht]
\center
\begin{subfigure}[t]{0.32\textwidth}
\centering
\includegraphics[width=\columnwidth]{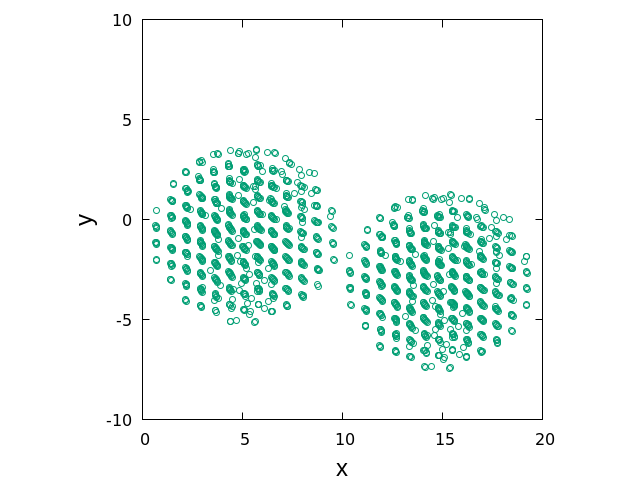}
\caption{$t = 5$}
\end{subfigure}
\begin{subfigure}[t]{0.32\textwidth}
\centering
\includegraphics[width=\columnwidth]{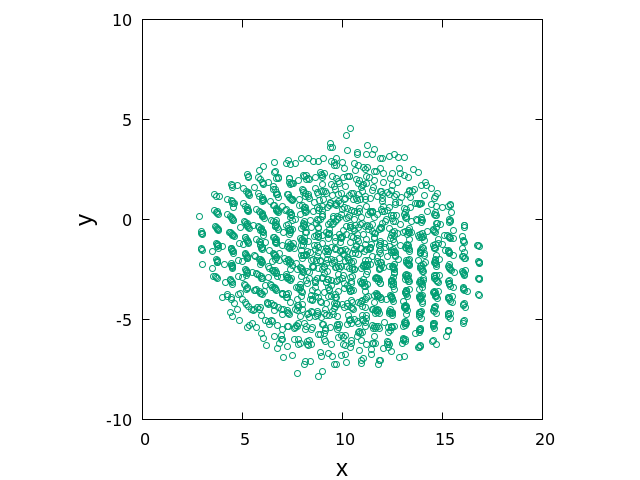}
\caption{$t = 8$}
\end{subfigure}
\begin{subfigure}[t]{0.32\textwidth}
\centering
\includegraphics[width=\columnwidth]{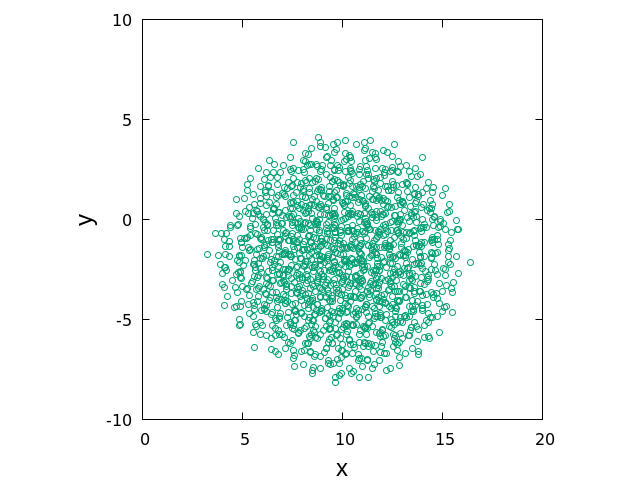}
\caption{$t = 40$}
\end{subfigure}
\caption{The example of a sticking collision for aggregates with the modified L-J potential for different time, $v_{\rm imp} = 2$. The  circles indicate individual particles.}
\label{fig:stick_mLJ}
\end{figure}
\begin{figure}[ht]
\center
\begin{subfigure}[t]{0.32\textwidth}
\centering
\includegraphics[width=\columnwidth]{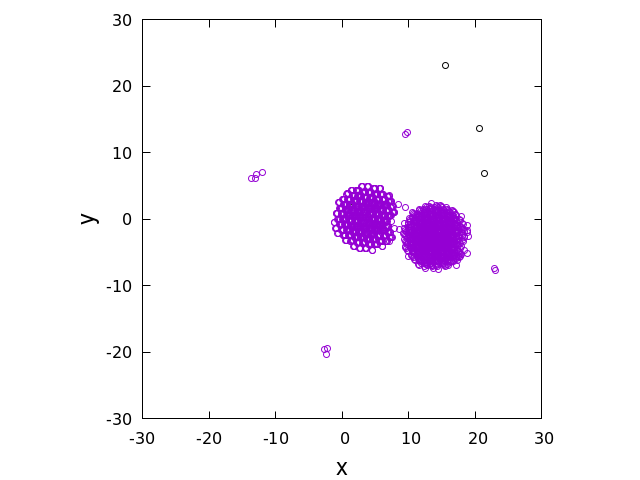}
\caption{ 
}
\label{fig:examplesa}
\end{subfigure}
\begin{subfigure}[t]{0.32\textwidth}
\centering
\includegraphics[width=\columnwidth]{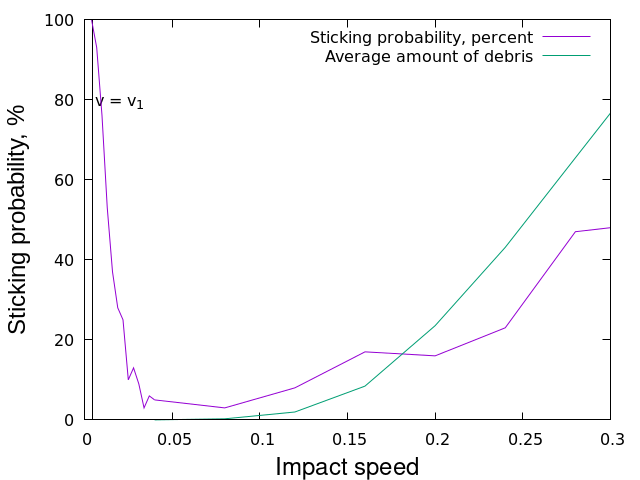}
\caption{
}
\label{fig:examplesb}
\end{subfigure}
\begin{subfigure}[t]{0.32\textwidth}
\centering
\includegraphics[width=\columnwidth]{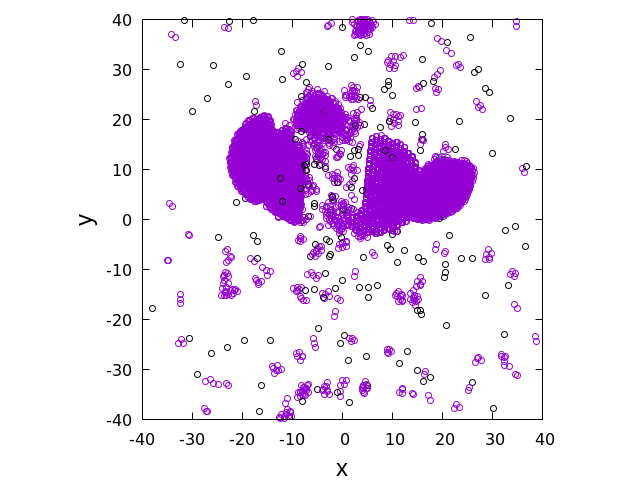}
\caption{
}
\label{fig:examplesc}
\end{subfigure}
\caption{Collisions with fragmentation and erosion for the JKR aggregates. Circles represent individual particles. Monomers are black, other aggregates are violet. (a) Bouncing with fragmentation, $v_{\rm imp} = 0.2$, $R = 5$. (b) Sticking probability for small impact speeds, based on 100 collisions for each speed,  $R = 5$. For $v_{\rm imp} <0.015$ the collisions are mostly sticking, for $0.015 <v_{\rm imp} <0.1 $ -- mostly bouncing and for $v_{\rm imp} > 0.1 $ are mostly bouncing with erosion. (c) Fragmentation for  $v_{\rm imp} = 1.4$,  $R=10$. }
\label{fig:examples}
\end{figure}

\Figl{fig:stick_mLJ} illustrates  a sticking collision for  the long-ranged modified L-J potential. As one can see from the figure,  after  the merging, the joint aggregate finally attains a spherical form. This happens due to the strong surface tension for this potential.  We do not show purely bouncing collision here (see, e.g. \cite{Saitoh} for the visualization of such impacts); instead, we present in \Fig{fig:examplesa} the bouncing collision with  erosion. Erosion may also accompany sticking, which is presented in a more  quantitative way in \Fig{fig:examplesb}, showing  a complicated non-monotonous dependence of the sticking probability on the impact velocity. The increase of the sticking probability with the increasing $v_{\rm imp}$ in the range of $0.1 < v_{\rm imp} < 0.3$ is accompanied by the increasing erosion. The latter  leads to the increasing contact  area and hence to the stronger interaction between the aggregates. To define a quantitative boundary between fragmentation and erosion, we use the criterion which follows from the universal scaling behavior of the shattering degree $S$,  depicted in \Fig{fig-mon}. Namely, we suggest defining a "developed" fragmentation as the process, obeying the scaling law \eqref{eq:S}. It starts from the threshold impact velocity $v_{\rm imp,0} $, where the scaling dependence \eqref{eq:S} yields $S=0$. In what follows we use   $v_{\rm imp,0} $ to demarcate the erosion from fragmentation. 

\begin{figure}[ht]
\center
\begin{subfigure}[t]{0.32\textwidth}
\centering
\includegraphics[width=\columnwidth]{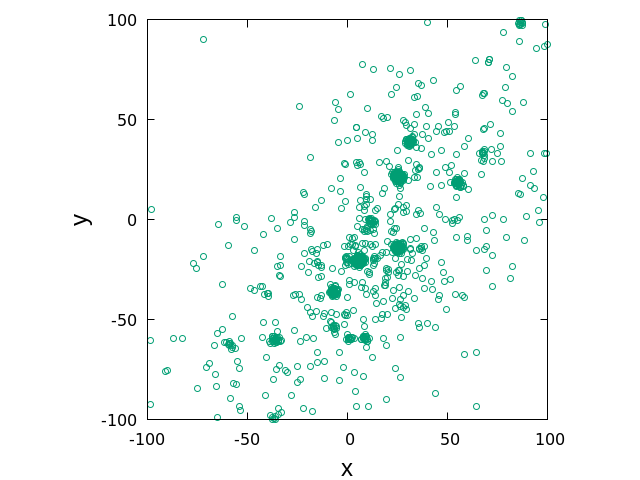}
\caption{ 
}
\label{fig:examples2a}
\end{subfigure}
\begin{subfigure}[t]{0.32\textwidth}
\centering
\includegraphics[width=\columnwidth]{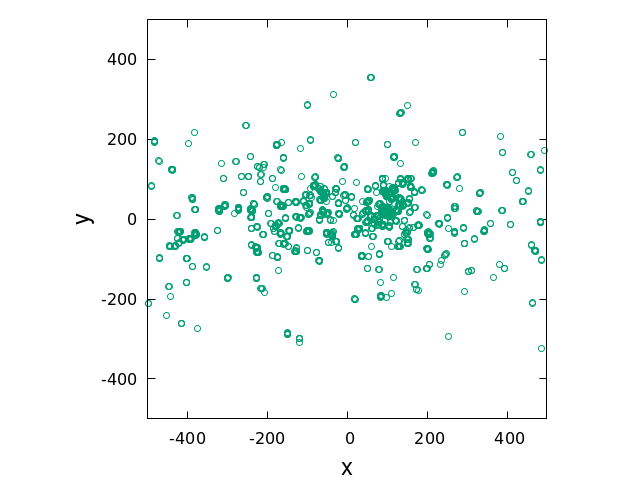}
\caption{
}
\label{fig:examples2b}
\end{subfigure}
\begin{subfigure}[t]{0.32\textwidth}
\centering
\includegraphics[width=\columnwidth]{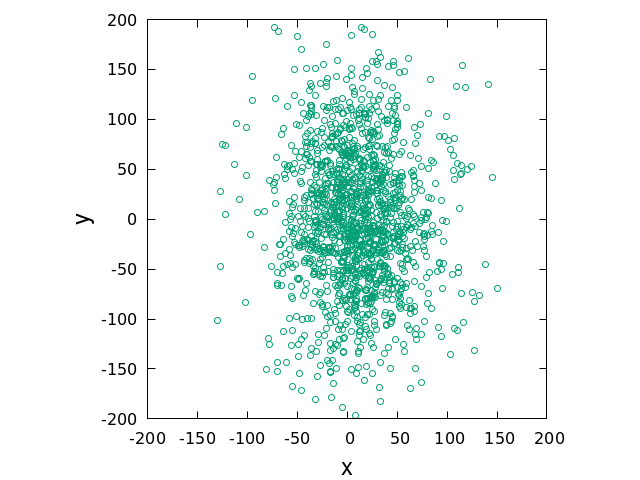}
\caption{
}
\label{fig:examples2c}
\end{subfigure}
\caption{(a) Fragmentation of the modified Lennard-Jones aggregates, $v_{\rm imp} = 10$, $R=5$. (b) Fragmentation of aggregates with the  Tersoff potential, $v_{\rm imp} = 200$, $R=2.87$ lattice units.  (c) Shattering of the Lennard-Jones aggregates, $v_{\rm imp} = 20$, $R=5$. Circles represent individual particles. }
\label{fig:examples2}
\end{figure}

\Figls{fig:examplesc} and \ref{fig:examples2a}-\ref{fig:examples2b} illustrate fragmentation of the JKR, modified L-J and Tersoff aggregates, while \ref{fig:examples2c} -- shattering of L-J aggregates. 

\begin{figure}[h!]
\begin{tikzpicture}[scale = 3.5]

\def\vs{0.8}

\large

\draw[->] (0,0+0.5*\vs) -- (4,0+0.5*\vs);
\draw (1pt,2.303/4*\vs+4*\vs) -- (-1pt,2.303/4*\vs+4*\vs) node[anchor=east] {10};
\draw (1pt,0+4*\vs) -- (-1pt,0+4*\vs) node[anchor=east] {1};
\draw (1pt,-2.303/4*\vs+4*\vs) -- (-1pt,-2.303/4*\vs+4*\vs) node[anchor=east] {0.1};
\draw (1pt,-4.605/4*\vs+4*\vs) -- (-1pt,-4.605/4*\vs+4*\vs) node[anchor=east] {0.01};
\draw (1pt,-6.908/4*\vs+4*\vs) -- (-1pt,-6.908/4*\vs+4*\vs) node[anchor=east] {0.001};
\draw (1pt,-9.210/4*\vs+4*\vs) -- (-1pt,-9.210/4*\vs+4*\vs) node[anchor=east] {0.0001};
\draw (1pt,-11.513/4*\vs+4*\vs) -- (-1pt,-11.513/4*\vs+4*\vs) node[anchor=east] {0.00001};
\draw (1pt,-13.815/4*\vs+4*\vs) -- (-1pt,-13.815/4*\vs+4*\vs) node[anchor=east] {0.000001};

\draw[decorate, decoration={brace, mirror, amplitude=3mm}, shorten <= 4pt, shorten >= 4pt] (0,0+0.5*\vs) -- node[below, align = center] {\\ JKR} (1,0+0.5*\vs);
\draw[decorate, decoration={brace, mirror, amplitude=3mm}, shorten <= 4pt, shorten >= 4pt] (1,0+0.5*\vs) -- node[below, align = center] {\\ Tersoff} (2,0+0.5*\vs);
\draw[decorate, decoration={brace, mirror, amplitude=3mm}, shorten <= 4pt, shorten >= 4pt] (2,0+0.5*\vs) -- node[below, align = center] {\\ L-J} (3,0+.5*\vs);
\draw[decorate, decoration={brace, mirror, amplitude=3mm}, shorten <= 4pt, shorten >= 4pt] (3,0+0.5*\vs) -- node[below, align = center] {\\ Mod. L-J} (4,0+0.5*\vs);




\fill [fill=green!20]
(2,0+0.5*\vs) -- (0,0+0.5*\vs) -- (0,5*\vs) -- (2,5*\vs) -- cycle;

\fill [fill=yellow!20]
(0,-11.787/4*\vs+4*\vs) -- (1,-11.787/4*\vs+4*\vs) -- (1,0+0.5*\vs) -- (0,0+0.5*\vs) -- cycle;
\fill [fill=yellow!20]
(1,-3.301/4*\vs+4*\vs) -- (2,-3.301/4*\vs+4*\vs) -- (2,-1.915/4*\vs+4*\vs) -- (1,-1.915/4*\vs+4*\vs) -- cycle;
\fill [fill=yellow!20]
(2,0+0.5*\vs) -- (4,0+0.5*\vs) -- (4,5*\vs) -- (2,5*\vs) -- cycle;

\fill [fill=orange!20]
(0,-5.992/4*\vs+4*\vs) -- (1,-5.992/4*\vs+4*\vs) -- (1,-1.915/4*\vs+4*\vs) -- (2,-1.915/4*\vs+4*\vs) -- (2,-3.943/4*\vs+4*\vs) -- (3,-3.943/4*\vs+4*\vs) -- (3,-1.679/4*\vs+4*\vs) -- (4,-1.679/4*\vs+4*\vs) -- (4,5*\vs) -- (0,5*\vs) -- cycle;

\fill [fill=red!20]
(0,-0.143/4*\vs+4*\vs) -- (1,-0.143/4*\vs+4*\vs) -- (1,-0.082/4*\vs+4*\vs) -- (2,-0.082/4*\vs+4*\vs) -- (2,-2.117/4*\vs+4*\vs) -- (3,-2.117/4*\vs+4*\vs) -- (3,-1.273/4*\vs+4*\vs) -- (4,-1.273/4*\vs+4*\vs) -- (4,5*\vs) -- (0,5*\vs) -- cycle;

\fill [fill=magenta!20]
(0,3.651/4*\vs+4*\vs) -- (1,3.651/4*\vs+4*\vs) -- (1,1.977/4*\vs+4*\vs) -- (2,1.977/4*\vs+4*\vs) -- (2,1.361/4*\vs+4*\vs) -- (3,1.361/4*\vs+4*\vs) -- (3,2.671/4*\vs+4*\vs) -- (4,2.671/4*\vs+4*\vs) -- (4,5*\vs) -- (0,5*\vs) -- cycle;

\draw[->] (0,0+0.5*\vs) -- (0,5*\vs) node[anchor=east] {$\frac{E_{kin}}{E_{con}}$};
\draw[dotted] (1,0+0.5*\vs) -- (1,5*\vs);
\draw[dotted] (2,0+0.5*\vs) -- (2,5*\vs);
\draw[dotted] (3,0+0.5*\vs) -- (3,5*\vs);

\draw[-] (0,3.651/4*\vs+4*\vs) -- (1,3.651/4*\vs+4*\vs);
\draw[-] (1,1.977/4*\vs+4*\vs) -- (2,1.977/4*\vs+4*\vs);
\draw[-] (2,1.361/4*\vs+4*\vs) -- (3,1.361/4*\vs+4*\vs);
\draw[-] (3,2.671/4*\vs+4*\vs) -- (4,2.671/4*\vs+4*\vs);

\draw[-] (1,3.651/4*\vs+4*\vs) -- (1,1.977/4*\vs+4*\vs);
\draw[-] (2,1.361/4*\vs+4*\vs) -- (2,1.977/4*\vs+4*\vs);
\draw[-] (3,2.671/4*\vs+4*\vs) -- (3,1.361/4*\vs+4*\vs);
\draw[-] (0,-0.143/4*\vs+4*\vs) -- (1,-0.143/4*\vs+4*\vs);
\draw[-] (1,-0.082/4*\vs+4*\vs) -- (2,-0.082/4*\vs+4*\vs);
\draw[-] (2,-2.117/4*\vs+4*\vs) -- (3,-2.117/4*\vs+4*\vs);
\draw[-] (3,-1.273/4*\vs+4*\vs) -- (4,-1.273/4*\vs+4*\vs);

\draw[-] (1,-0.082/4*\vs+4*\vs) -- (1,-0.143/4*\vs+4*\vs);
\draw[-] (2,-2.117/4*\vs+4*\vs) -- (2,-0.082/4*\vs+4*\vs);
\draw[-] (3,-1.273/4*\vs+4*\vs) -- (3,-2.117/4*\vs+4*\vs);
\draw[-] (0,-5.992/4*\vs+4*\vs) -- (1,-5.992/4*\vs+4*\vs);
\draw[-] (1,-1.915/4*\vs+4*\vs) -- (2,-1.915/4*\vs+4*\vs);
\draw[-] (2,-3.943/4*\vs+4*\vs) -- (3,-3.943/4*\vs+4*\vs);
\draw[-] (3,-1.679/4*\vs+4*\vs) -- (4,-1.679/4*\vs+4*\vs);
\draw[-] (0,-11.787/4*\vs+4*\vs) -- (1,-11.787/4*\vs+4*\vs);
\draw[-] (1,-3.301/4*\vs+4*\vs) -- (2,-3.301/4*\vs+4*\vs);

\draw[-] (1,0+0.5*\vs) -- (1,-11.787/4*\vs+4*\vs);
\draw[-] (2,0+0.5*\vs) -- (2,-3.943/4*\vs+4*\vs);

\draw[-] (1,-5.992/4*\vs+4*\vs) -- (1,-1.915/4*\vs+4*\vs);
\draw[-] (2,-3.943/4*\vs+4*\vs) -- (2,-2.117/4*\vs+4*\vs);
\draw[-] (3,-3.943/4*\vs+4*\vs) -- (3,-2.117/4*\vs+4*\vs);

\node[align = center] at (2,4.8*\vs) {\huge Shattering};
\node[align = center] at (2,4.2*\vs) {\huge Fragmentation};

\node[align = center] at (1,3.75*\vs) {\Large Erosion};
\node[align = center] at (2.5,3.25*\vs) {Erosion};
\node[align = center] at (3.5,3.64*\vs) {\small Erosion};

\node[align = center] at (1.5,3.35*\vs) {Sticking};
\node[align = center] at (0.5,0.75*\vs) {Sticking};
\node[align = center] at (3,1.7*\vs) {\huge Sticking};

\node[align = center] at (1,1.7*\vs) {\huge Bouncing};


\end{tikzpicture}
\caption{
Collision type diagram for four potentials. Separation lines are obtained as described in the text. The  vertical axis shows the ratio of the kinetic energy $E_{\rm kin}$ and  the total potential energy of the  aggregates $E_{\rm pot}$. The horizontal axis indicates the potentials studied. 
}
\label{fig-diag}
\end{figure}
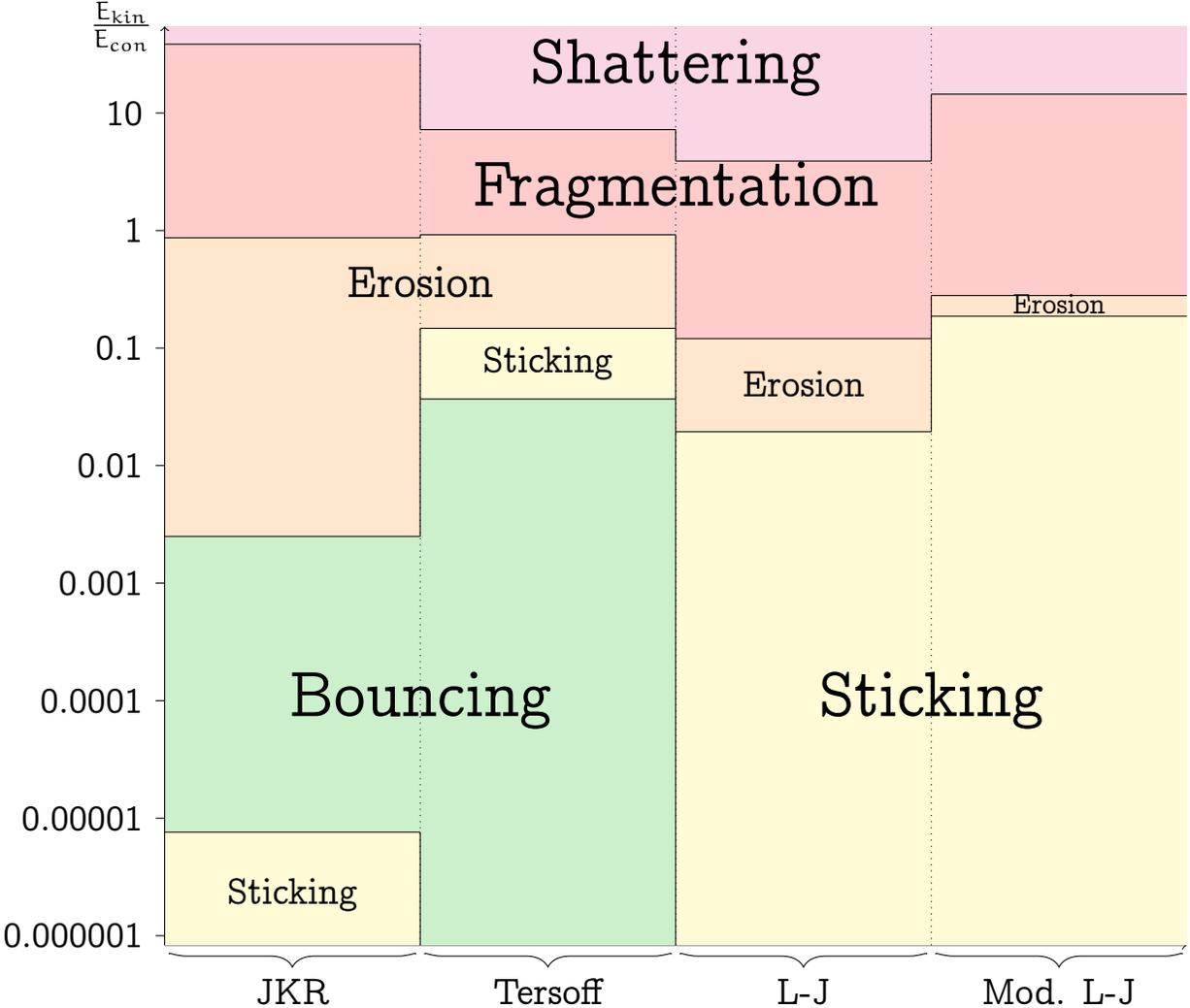

All the collision regimes for the  four potentials are schematically presented in \Fig{fig-diag}. Note that some boundaries are conditional, as mixed regimes are possible (see the discussion above). Obviously, the collision outcome is determined by the interplay between the kinetic energy, $E_{\rm kin}= 2 N (v_{\rm imp}/2)^2/2$, and the total potential energy of two aggregates $E_{\rm pot}$. The former drives fragmentation/shattering of the clusters; the latter -- preserves their integrity and favors sticking. Hence the ratio $E_{\rm kin}/E_{\rm pot}$ is a natural measure of collision strength. As it may  be seen from \Fig{fig-diag} the collision regime drastically depends on the nature of a potential. For instance, aggregates with both L-J and modified L-J potentials do not demonstrate bouncing collisions. This may be explained by the relatively flat minimum of the potential well, where the monomers in the aggregates are placed (more pronounced for the modified L-J).  In this case, plastic deformation of the material with the rearrangement of atoms is plausible; it suppresses the elastic response and prevents bouncing. The Tersoff potential is more short-ranged than the L-J potential and thus responds elastically up to relatively high deformations (impact speeds), which results in bouncing. At still higher impact speeds, the accompanying erosion leads to the increasing contact area, which causes sticking. The JKR interactions are the  most short-ranged. Sticking at small impact velocities is associated with the irreversible formation of contacts between constituents of two aggregates. As long as the collision energy is smaller than that of the formed contacts, the aggregates stick; this is similar to the sticking of JKR monomers analyzed in \cite{Brilliantov2007}. The sticking regime, in this case, is rather limited, as the JKR aggregates are very fragile and can hardly sustain plastic deformation. As a result, with the increasing collision energy, sticking transforms into bouncing and then to erosion (or bouncing with erosion). For high collision energy,  the behavior for all potentials is similar: As the ratio $E_{\rm kin}/E_{\rm pot}$ increases, erosion transforms into fragmentation and the latter one into shattering.

\section{Qualitative theory }

As has been already reported  in \cite{basis} and confirmed in our study,  at least for two potentials, \Fig{speed-fig}, the average kinetic energy of a  fragment scales with  its mass as a power law, $E_{\rm kin,f} \sim m^{1/3}$. Here we discuss the theoretical justification of this dependence. 

We analyze the case of "developed" fragmentation, that is, the debris size ranges from monomers to largest fragments, which are significantly smaller than a whole aggregate. Moreover, we consider the fragments which comprise many  monomers. We assume that all fragments arise in a single event of the duration $\tau$, which generates uniform internal tension, driving the fragmentation.  Then the force acting on a fragment of size $k$ would be proportional to its cross-section, which scales as $\sim r_0^2 k^{2/3}$ (a fragment of size $k$ is comprised of $k$ monomers of size $\sim r_0$ and mass $m_0$). Hence the acceleration of this fragment with the mass $m_0k$ scales as $r_0^2k^{2/3}/(m_0k) \sim (r_0^2/m_0)k^{-1/3}$. Since we assume that the duration of the force action is the same for all fragments, we conclude that their velocity would scale with size as $v  \sim \tau (r_0^2/m_0)k^{-1/3} \sim k^{-1/3}$. Correspondingly, 

\begin{equation}
\label{eq:E_m}
 E_{\rm kin,f} = \frac12 mv^2 \sim m \, m^{-2/3} \sim m^{1/3}.    
\end{equation}
While the scaling \eqref{eq:E_m} holds for a wide range of fragments masses only for some potentials, the scaling depicted in \Fig{fig-mon} is more enigmatic, as it  seems to be universal for all studied potentials. Below we try to give a plausible explanation for this behavior.

Let $n_{\rm ag}(v)= 2N-n_1$ be the total number of monomers  which belong to aggregates  with  the size $k \geq 2$, arising  at a collision with the impact velocity $v=v_{\rm imp}$. With the increasing $v$, some of them will further  split, producing more monomers. Hence the decay rate of $n_{\rm ag}(v)$ with the growing $v$ will be proportional to $n_{\rm ag}$ itself, as the number of produced monomers is proportional to the number of monomers in all clusters. As the monomer splitting is a random process, we assume that it is similar to a  Poisson process and obeys its rules. Then, the probability that the monomer will separate from a fragment with size $k\geq 2$, per an increment of the impact velocity $d v_{\rm imp}$,  may be estimated as for a Poisson process, as $b \,d v_{\rm imp}$, with a constant $b$. Hence $d  n_{\rm ag} \sim - b \,d v_{\rm imp}\, n_{\rm ag}$, or

\begin{equation}
\label{eq:nag}
\frac{d n_{\rm ag}}{d v_{\rm imp}} \sim - b n_{\rm ag}.
\end{equation}
Integrating Eq. \eqref{eq:nag} yields, $n_{\rm ag} \sim e^{-b v_{\rm imp}}$. Since $1-S \sim n_{\rm ag} $, see Eq. \eqref{eq:S}, we obtain the observed dependence \eqref{eq:S1}.  

\section{Conclusion}
We undertake a comprehensive study of collision fragmentation of aggregates comprised of particles with different interaction potentials. We analyzed four potentials -- the Lennard-Jones  (L-J), Tersoff ,  modified L-J potential and the one corresponding to the Johnson-Kendall-Roberts (JKR) model. The first two potentials are used to model inter-atomic interactions and are middle-ranged. The JKR model refers to the interactions of macroscopic adhesive bodies and is extremely short-ranged. Finally, the  modified L-J potential was constructed as a long-ranged potential to investigate the impact of the interaction range of the fragmentation process. Thus we explore a set of potentials of very different ranges and nature. 

We perform molecular dynamics (MD) simulations  and confirm previously reported power-law dependence for the fragments mass distribution and for the dependence of the  fragments after-collisional velocity  on their  mass. We also investigate the  angular distribution  of the directions of the fragments velocities and quantify it in terms of the Legendre polynomial series. 

We reveal a new surprising scaling for the amount of monomer  production at a collision and define a new fragmentation characteristic -- the shattering degree $S$. We observe that $1-S$ obeys a universal exponential dependence on the impact velocity $v_{\rm imp}$ for a wide range of $v_{\rm imp}$ and for all four potentials. 

We perform a classification of collisions by their outcome and construct the corresponding phase diagram. It illustrates different collision types for different potentials. Namely, we distinguish five different types of collisions: sticking, bouncing, erosive, shattering impacts and these with  fragmentation. We also observe mixed types of collisions. Using the newly revealed  scaling  for the monomer production, we propose a quantitative criterion discriminating erosion and fragmentation. We observe that although all studied potentials demonstrate similar scaling, the location of the different  collision regimes on the phase diagram drastically differs and depends on their interaction range. 
Finally, to explain the observed scaling dependencies, we develop a qualitative theory. 

Although we restrict our study by head-on collisions, we believe that the oblique collisions will demonstrate similar features. Moreover, we also deem that the main qualitative laws observed for small clusters are generic and will persist for large aggregates. 

\vskip 0.3cm
The study was supported by a grant from the Russian Science Foundation No. 21-11-00363, https://rscf.ru/project/21-11-00363/; AO also acknowledges RFBR project No. 20-31-90022. Zhores supercomputer of Skolkovo Institute of Science and Technology \cite{zhores} has been used in the present research.

\appendix{}

\section{Tersoff potential parameters}

Tersoff potentials \cite{ters1989, ters1988} are based on the exponential two-particle potential, where the total interaction energy  between atoms $i$ and $j$ reads, 
\[
  E_{ij} = \left( A \exp \left(- \lambda_1 r_{ij} \right)  - B_{ij} \exp \left( - \lambda_2 r_{ij} \right) \right) f_C ( r_{ij} ),
\]
with the coefficient $B_{ij}$, depending on other particles in the interaction sphere and the angles between the bonds. Here $r_{ij}$ is the distance between atoms $i$ and $j$. The exact form of the dependence of $B_{ij}$ on other particles is
\[
\begin{aligned}
  B_{ij} & = B \left( 1 + \beta^n \zeta_{ij}^n \right)^{-1/(2n)}, \\
  \zeta_{ij} & = \sum\limits_{k \ne i, j} f_C ( r_{ik} ) g \left( \theta_{ijk} \right), \\
  g \left( \theta \right) & = 1 + \frac{c^2}{d^2} - \frac{c^2}{d^2 + \left( \cos{\theta_0} - \cos{\theta} \right)^2},
\end{aligned}
\]
where $\theta_{ijk}$ is the angle in radians between the bonds $ij$ and $ik$.
Function $f_c(r)$ describes a smooth cutoff from distance $R$ to $R+D$:
\[
{f_C}\left( r \right) = \begin{cases}
   {1} & {r \leqslant R - D}  \\ 
   {\tfrac{1}{2} - \tfrac{1}{2}\sin \left( {\tfrac{\pi }{2}\left( {r - R} \right)/D} \right)} & {R - D < r \leqslant R + D}  \\ 
   {0} & {r > R + D} 
\end{cases} 
\]

In simulations, we use the following parameters:
\[
\begin{aligned}
  c & = 38049, \\
  d & = 4.3484, \\
  \cos{\theta_0} & = -0.57058, \\
  n & = 0.72751, \\
  \beta & = 0.00000015724, \\
  \lambda_2 & = 2.2119 A^{-1}, \\
  B & = 346.7 eV, \\
  R & = 1.95 A, \\
  D & = 0.15 A, \\
  \lambda_1 & = 3.4879 A^{-1}, \\
  A & = 1393.6 eV.
\end{aligned}
\]

In the figures, we plot the speed in angstroms per picosecond, which is a standard unit in LAMMPS, $1 A/ps = 100 m/s$ (time unit is 1 picosecond). Sphere radius in figures is given in lattice units (1 lattice unit for diamond lattice is 3.57 angstroms). The mass unit in figures corresponds to the mass of one carbon atom $m_C = 12.0107 g/mol$, as used in simulations.

\section{Numerical algorithm and more simulation detail}\label{alg-sec}
To keep our algorithm as simple as possible, we used the leap-frog scheme \cite{leap-frog}.
At $k$-th time step $t_k = k \Delta t$ it yields: 
\begin{equation}\label{eq-leap}
\begin{aligned}
\vec v_i(t_k + \Delta t/2) & = \vec v_i(t_k - \Delta t / 2) + \mathop \sum\limits_{j} \vec F_{ij}(t_k) \\ 
\vec x_i(t_k + \Delta t) & = \vec x_i(t_k) + \vec v_i(t_k + \Delta t/2).
\end{aligned}
\end{equation}
To use this scheme, we need to determine the initial condition for the speed at $\vec v_i(\Delta t/2)$. We did it as follows. Since each aggregate was at zero temperature and they were far apart from each other, we concluded that the force acting on each particle $i$ vanished  and remained zero for a sufficiently long time. Correspondingly, the speeds of the particles did not initially change. 
\begin{equation}\label{eq-vt}
  \mathop \sum\limits_{j} \vec F_{ij}(0) = 0, \qquad \vec v_i (\Delta t/2) = \vec v_i (0), 
\end{equation}
Hence, according to Eq. (\ref{eq-vt}) we could use the initial velocities of the  whole aggregates as the starting values for $\vec v_i (0)$.\footnote{Using $\vec v_i'(t) = \vec v_i(t + \Delta t/2)$ instead of $v_i$ in equation (\ref{eq-leap}) is equivalent to the application  of the Euler scheme, instead of the leap-frog scheme, which surprisingly, implies that in our case the Euler scheme corresponds to the second order approximation.}

\begin{figure}[h!]
\center
\begin{subfigure}[t]{0.49\textwidth}
\centering
\includegraphics[width=\columnwidth]{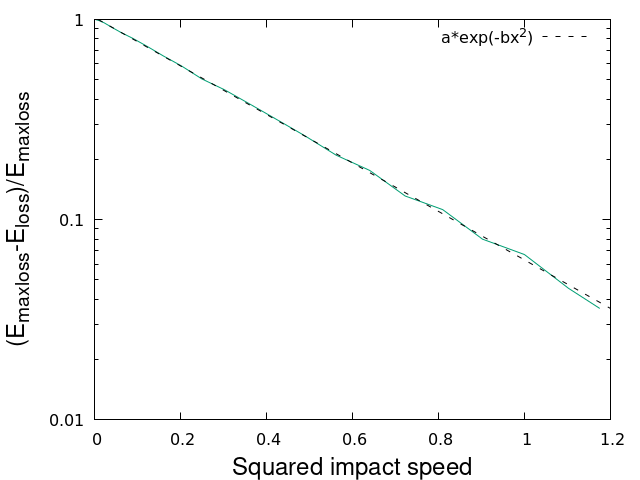}
\caption{Johnson-Kendall-Roberts (JKR) \\ interactions,  $R = 5$, \\ averaging is  over 100 collisions. \\ $a = 1.025 \pm 0.007, b = 2.790 \pm 0.016.$}
\vspace*{3mm}
\end{subfigure}
\begin{subfigure}[t]{0.49\textwidth}
\centering
\includegraphics[width=\columnwidth]{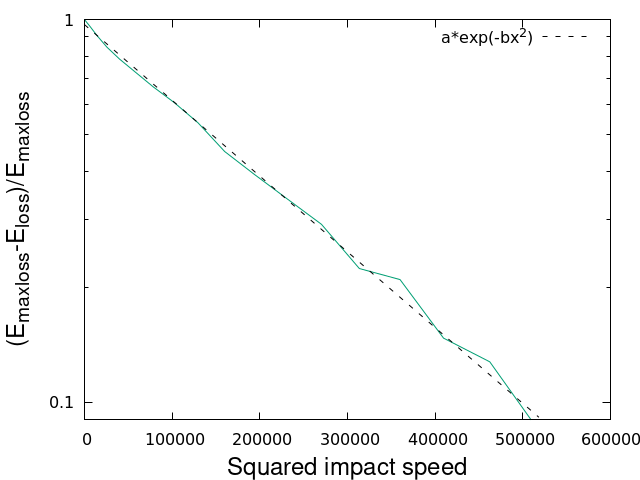}
\caption{Tersoff carbon potential,  $R = 2.87$ lattice units, \\ averaging is over 10 collisions. \\ $a = 0.970 \pm 0.015, b = (4.556 \pm 0.006) \cdot 10^{-6}.$}
\vspace*{3mm}
\end{subfigure}
\begin{subfigure}[t]{0.49\textwidth}
\centering
\includegraphics[width=\columnwidth]{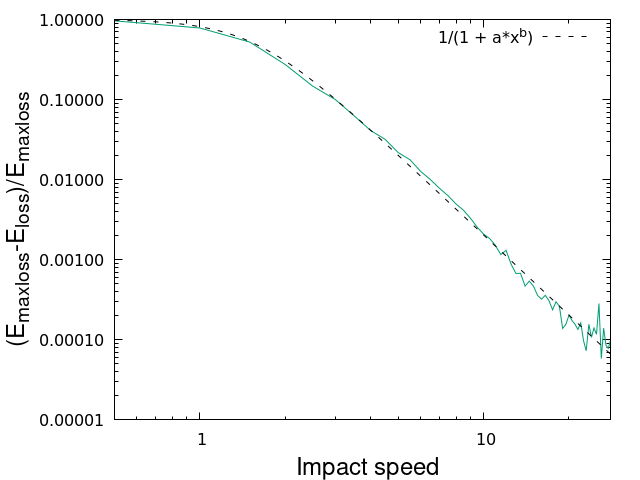}
\caption{Lennard-Jones potential,  $R = 5$, \\ averaging is over 10 collisions. \\ $a = 0.23 \pm 0.04, b = 3.33 \pm 0.06.$}
\end{subfigure}
\begin{subfigure}[t]{0.49\textwidth}
\centering
\includegraphics[width=\columnwidth]{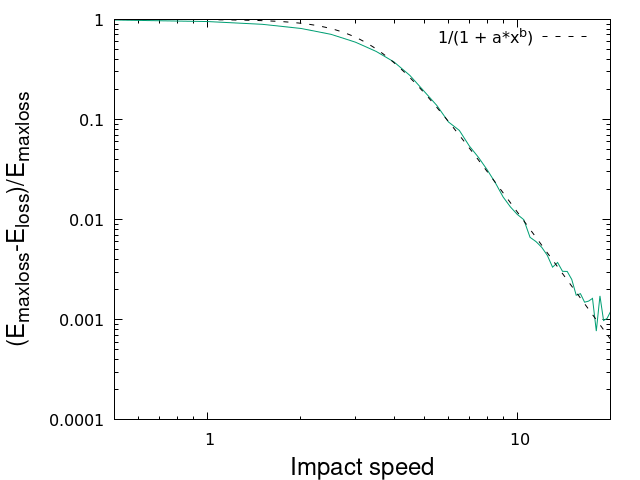}
\caption{Modified Lennard-Jones potential,  $R = 5$, \\ averaging is over 10 collisions. \\ $a = (4.8 \pm 0.9) \cdot 10^{-3}, b = 4.24 \pm 0.08.$}
\end{subfigure}
\caption{The ratio $(E_{\rm maxloss}-E_{\rm loss})/E_{\rm maxloss}$, where $E_{\rm loss}$ is the kinetic energy loss at a collision  and $E_{\rm maxloss}$ is the maximum possible kinetic energy loss. For all potentials, except JKR, the  maximum possible loss of the kinetic energy is equal to the total initial potential energy $E_{\rm maxloss} = \left| E_{\rm pot} \right|$. For the JKR $E_{\rm maxloss}$ also includes the maximum energy loss, associated with the irreversible  formation and breaking of monomers contacts.}
\label{fig-eloss}
\end{figure}

All force calculations were parallelized with the OpenMP. For the JKR, we have used the Verlett method \cite{leap-frog} for the fast update of the neighbor lists. We have used the following time steps: $0.02$ for L-J, $0.01/\max(1,v)$ for the modified L-J, $0.002$ for JKR, and $0.0005 ps$ for the Tersoff potential.

Also, note that we used $R = 10$ for the JKR force, while for other potentials we used a smaller radius. Larger radius for the JKR was  required due to a scarcity of  large fragments for these short-range interactions (see \Fig{fig:conc-jkr}). Hence we needed  larger colliding aggregates  for accurate detection of the expected power-laws. Moreover, due to the nature of the JKR potential (we used it in the form of the inter-particle force), it has an additional randomness in the amount of possible energy loss. As a result, much more experiments for the JKR aggregates were  needed when the role of dissipation was high. Therefore, we used $R = 5$ (the same as for other potentials) for the JKR in \Figs{fig:examplesa}-\ref{fig:examplesb} and in \Fig{fig-eloss}. To get similar maximum displacement and keep potential energy per bond the same as for $R = 10$ (which guarantees that the same time step may be used and fragmentation occurs at the same impact speeds), we  re-scaled the parameters as follows: $D_0 \sim N^{-1}$ and $\varkappa \sim N^{-0.3}$, where $N$ is the number of monomers in each of the initial clusters, leading to $D_0 = 8.81 \cdot 10^{-5}$ and $\varkappa = 9.68 \cdot 10^{-3}$ for $R = 5$ with $N = 757$ monomers.

For completeness, we report here the analysis for the energy loss at the disruptive collisions. As the impact velocity increases, the energy loss should converge to the maximal possible value, corresponding to all broken bonds.  As \Fig{fig-eloss} shows, this convergence is not universal but demonstrates  two distinct scalings: for the L-J and modified L-J aggregates, the remaining bond energy scales polynomially. In contrast, for the JKR and Tersoff aggregates, it scales as $\exp(-bv_{\rm imp}^2)$.

\bibliographystyle{unsrt}

\end{document}